\documentclass[11pt]{article}
\pdfoutput=1
\usepackage{jheppub}
\usepackage{graphicx}
\usepackage{amssymb}
\usepackage{amsmath,amssymb}
\usepackage{slashed}
\usepackage{hyperref}
\usepackage{caption}
\usepackage{xcolor}
\usepackage{dsfont}
\usepackage{verbatim}
\usepackage{subfig}
\usepackage[usestackEOL]{stackengine}

\newcommand\beq{\begin{equation}}
\newcommand\eeq{\end{equation}}
\newcommand\be{\begin{equation}}
\newcommand\ee{\end{equation}}

\newcommand\ra{\rightarrow}

\title{Reduced conformal symmetry}

\preprint{\today}

\author[a]{Andreas Karch,}
\author[a]{Amir Raz}
\affiliation[a]{University of Texas, Austin, Physics Department, Austin TX 78712, USA}

\emailAdd{karcha@utexas.edu}
\emailAdd{araz@utexas.edu}
\abstract{We construct field theories in $2+1$ dimensions with multiple conformal symmetries acting on only one of the spatial directions. These can be considered a conformal extension to ``subsystem scale invariances", borrowing the language often used for fractons.
}

\begin{document}
\maketitle

\strutlongstacks{T}

\section{Introduction}

The study of new and exotic phases of matter has seen an resurgence in recent years with the discovery of fractons (see \cite{Nandkishore_2019,Pretko_2020} for a comprehensive review). These systems have particle excitations with restricted mobility, and large ground state degeneracy; phenomena that do not fit nicely into the standard framework of continuum quantum field theory. 

Continuum quantum field theories are typically described as a renormalization group (RG) flow between fixed points, or conformal field theories. This gives us a universal description of the low-energy physics without knowing the exact microscopic details of the model. From generic RG arguments we expect such fixed points to have some sort of scaling symmetry and translational symmetries, though as rotations are explicitly broken in fractonic models the fixed points cannot have the full $2+1$ dimensional conformal symmetry. Our goal in this paper is to try and understand the space-time symmetries of fractonic RG fixed points.

Explicitly, we shall study the space-time symmetries of four ``free" (i.e. quadratic) field theories in $2+1$ dimensions, inspired by the continuum fracton models in \cite{Seiberg:2020bhn,Pretko_2018}. We show that these four models have sub-dimensional conformal symmetry in the two primary $1+1$ planes. This is more expansive than the subconformal symmetry observed in \cite{Slagle_2017}, or the anisotropic scaling considered in \cite{Gromov2019}. Furthermore, we show that the two independent conformal symmetries in general do not commute with each other, and so the full space-time symmetry algebra may be much larger. 

The restriction to free field theories may appear very limiting, but we should note that the free actions in \cite{Seiberg:2020bhn} include all relevant and marginal terms consistent with symmetries, so they are the most general low energy descriptions for the corresponding lattice models, even if the lattice model itself has short range interactions. Even though it of course would be interesting to generalize our construction to interacting field theories, the fact that the continuum field theory description tells us that, given the symmetry, the entire physics of lattice models is forced to be described by a free theory at long distances is one of its strengths, not a weakness. Among the models we consider, we will find some where the conformal symmetry algebra alone doesn't close, hinting at an enlarged symmetry with an infinite number of generators. It is possible that, in these cases, the subsystem symmetry we uncover alone is strong enough to disallow any interacting theories, though at least in one of these cases we propose a scheme to obtain interacting theories.

The four models we consider in this paper are the complex version of the scalar field theory from \cite{Seiberg:2020bhn}, as well as three simple generalizations of this theory. Similar scalar models were also considered in \cite{Gromov2019,Gromov_2020,radicevic2019systematic,Pretko_2018}. All the models we consider contain a single complex scalar field $\phi$ in $2+1$ dimensions. We study both a relativistic time derivative term in the action, as well as a non-relativistic single time derivative term. The spatial derivative term in the action is taken to be the four derivative term $\left|\partial_x \partial_y \phi\right|^2$ from \cite{Seiberg:2020bhn}, or the similar two derivative term $\phi^*\partial_x \partial_y \phi$. These terms are the lowest order terms consistent with the symmetries of a cubic or a rectangular lattice respectively. A summary of the models we consider is given in table \ref{tab:model_summary}. We shall refer to the models by their number of time and space derivatives, $(n,m)$, where $n$ is the number of time derivatives and $m$ the number of spatial derivatives. 

The exact nature of the complex scalar field $\phi$ is not crucial to the analysis of the sub-dimensional conformal symmetries. When considering a relativistic time derivative term one can take $\phi$ to be real, or even compact, and the results should still hold. Similarly $\phi$ can be a fermionic field in the theories with the non-relativstic time derivative term. The nature of $\phi$ does effect the spectrum and stability of the different models.

\begin{table}[]
    \centering
    \begin{tabular}{c|c|c|c}
        Model & Lagrangian & \Centerstack{Sub-dimensional \\ Conformal symmetry} & Extended symmetry \\
        \hline
        $(2,4)$ &  $\left| \partial_t \phi\right|^2 + \left|\partial_x \partial_y \phi\right|^2 $ & Full relativistic $2d$ & Infinite Dimensional, unknown \\
        $(2,2)$ &  $\left| \partial_t \phi\right|^2 + \phi^*  \partial_x \partial_y \phi $ & Schr\"odinger $1+1$ & $SO(3,2)$\\ 
        $(1,4)$ &  $i \phi^*  \partial_t \phi + \left|\partial_x \partial_y \phi\right|^2 $ & Schr\"odinger $1+1$ & Large, unknown \\ 
        $(1,2)$ &  $i \phi^*  \partial_t \phi + \phi^*  \partial_x \partial_y \phi $ & Holomorphic $2d$ & Schr\"odinger $1+(1,1)$
    \end{tabular}
    \caption{The four models we consider, and a summary of their sub-dimensional and extended space time symmetries.}
    \label{tab:model_summary}
\end{table}

Finally, we note that these models have the momentum dipole symmetry from \cite{Seiberg:2020bhn} given by the shift $\phi \rightarrow \phi + f(x) +g(y)$ for arbitrary functions $f$ and $g$. This leads to sub-dimensional conservation laws, as in \cite{Seiberg:2020bhn,Gromov_2020}, though the analysis of these global symmetries and their consequences is not the focus of this paper. One can deform these models by adding terms in the Lagrangian that break this symmetry down to a subgroup where $f(x)$ and $g(x)$ are constrained to be linear function \cite{Seiberg:2020bhn,Pretko_2018}. We will analyze the interplay between these deformations and conformal symmetry for the (2,4) model of section \ref{sec:2_4_model}, and the (1,4) model of section \ref{sec:1_4_model}.

The paper is organized as follows. We study the sub-dimensional conformal symmetries of each of these models individually in sections \ref{sec:2_4_model} through \ref{sec:1_4_model}. Then we present a summary of our results along with a discussion of future directions in section \ref{sec:summary}.

\section{The $(2,4)$ model} \label{sec:2_4_model}

We shall start our analysis by looking at a simple example model based on the scalar field theory of the X--Y plaquette model from \cite{Seiberg:2020bhn}. The theory consists of a complex scalar field $\phi$ in $2+1$ dimensions, with the Lagrangian 
\begin{equation}
\label{eq:ssmodel}
    \mathcal{L}  = \left|\partial_t \phi\right|^2 + \left|\partial_x \partial_y \phi \right|^2.
\end{equation}

This theory has two non-commuting sub-dimensional conformal symmetries. To make this evident we can expand $\phi$ in Fourier modes in the $y$ direction as
\begin{equation} \label{eq:phi_of_ky}
    \phi(t,x,y) = \int \frac{d k_y}{\sqrt{2\pi}} e^{iy k_y} \phi_{k_y}(t,x),
\end{equation}
and then we get the Lagrangian of a free relativistic particle 
\begin{equation} \label{eq:lag_rel_ky}
    \mathcal{L}  = \left|\partial_t \phi_{k_y} \right|^2 + k_y^2 \left|\partial_x \phi_{k_y} \right|^2.
\end{equation}

As this is simply a free boson, it has a $1+1$ dimensional conformal symmetry in the $x-t$ plane. For example, we can define the holomorphic coordinates 
\begin{equation}
    z_x = t + i\frac{x}{k_y}, \qquad \qquad \bar{z}_x = t-i\frac{x}{k_y}, 
\end{equation}
with the conformal generators taking the form
\begin{equation}
    L_n^x = - z_x^{n+1} \partial_x, \qquad \qquad 
    \bar{L}_n^x = - \bar{z}_x \bar{\partial}_x.
\end{equation}
These obey the holomorphic and anti-holomorphic Witt algebra's commutation relations. 

We can write the generators of the global conformal transformations explicitly as
\begin{equation}
    \begin{aligned}
    & H = \partial_t, \qquad \qquad && P_x = k_y \partial_x, \qquad \qquad
    && D_x = t\partial_t + x\partial_x,\\
    & M_x = \frac{1}{k_y} x\partial_t -k_y t\partial_x, \quad 
    && K_x = \frac{1}{k_y} x^2 \partial_x - k_y t^2 \partial_x +\frac{2}{k_y}tx\partial_t, \quad  
    && C_x = t^2\partial_t -\frac{1}{k_y^2} x^2\partial_t +2xt\partial_x.
    \end{aligned}
\end{equation}

We initially made a choice to expand $\phi$ in Fourier modes in the $y$ direction. We could have instead chosen to expand $\phi$ in terms of modes in $x$ direction, This would result in another free theory, where a different $1+1$ conformal symmetry would be evident. This algebra would be generated by the Witt algebra of the holomorphic coordinates 
\begin{equation}
     z_y = t + i\frac{y}{k_x}, \qquad \qquad \bar{z}_y = t-i\frac{y}{k_x}. 
\end{equation}

These two independent conformal algebras do not commute with each other. The commutation relations between the different conformal operators can be computed using the observation that the momentum modes, $k_i$, and coordinates, $x_i$, obey the canonical commutation relations. As far as we can tell the combined algebra that contains both conformal algebras does not seem to have any known structure, and the commutation relations do not seem to close on themselves. This is the case even when considering the algebra of the global conformal symmetry and not the full Witt algebra. We note that there is some freedom in choosing the holomorphic coordinate $z_x$ (or $z_y$) as we can scale it by any factor of $k_y$ (respectively $k_x$) as $k_y$ is treated as a parameter of the subsystem, but there does not seem to be a preferred scaling that solves the aforementioned problems.

Unfortunately these algebras do not have a simple representation in coordinate space. Even when considering the subset of global conformal symmetries, only translations and scaling have a nice representation in terms of the original coordinates. This hinders our ability to study this symmetry and it's implications. The existence of an infinite number of generators, along with the fact that this free action includes all relevant and marginal terms consistent with the global symmetries \cite{Seiberg:2020bhn}, may be enough to forbid any interacting theories with the same sub-system conformal symmetry. 

We can however ask what happens to the subsystem symmetry when we deform the model by a term that breaks the global dipole symmetry $\phi \rightarrow \phi + f(x) +g(y)$. In particular this model allows for interesting deformations under which this symmetry is broken to the subgroup generated by linear functions $f(x)$ and $g(x)$ as anticipated in \cite{Pretko_2018}. We'd like to analyze how this deformation in the particular example of the $(2,4)$ model affects the conformal symmetry that was the main focus of our work.

The deformation of interest that needs to be added to the Lagrangian of eq. \eqref{eq:ssmodel} reads \cite{Seiberg:2020bhn}
\begin{equation}
    \Delta {\cal L} = - \frac{1}{2 \mu'} \left [ (\partial_x^2 \phi)^2 + (\partial_y^2 \phi)^2 \right ].
\end{equation}
In momentum space this reads
\begin{equation}
    \Delta {\cal L} = - \frac{1}{2 \mu'} (k_x^4 + k_y^4) \phi^2.
\end{equation}
If we try to apply the same tricks as above, that is we treat $k_y$ as a fixed parameter when studying the symmetry of the 1+1 dimensional system formed by the $x$ and $t$ coordinates, we immediately note that the conformal subsystem symmetry acting on $x$ and $t$ alone is broken. $k_y^4$ appears as a mass term, so instead of studying a massless 1+1 dimensional scalar with its full conformal symmetry, we study a massive scalar which doesn't even preserve the scale invariance. So most of the symmetry we previously identified is broken by this deformation. What does remain intact is the diagonal scale invariance generated by
\begin{equation}
    D=D_x + D_y = 2 t\partial_t + x \partial_x + y \partial_y
\end{equation}
which is the standard non-relativistic $z=2$ scale invariance under which time gets rescaled twice as much as space. From the 1+1 dimensional point of view, we basically restore scale invariance by treating the $k_y^4$ mass term as spurion, that is we let it transform under the scale transformation in order to restore the symmetry.

\section{The $(2,2)$ model} \label{sec:2_2_model}

Motivated by the sub-dimensional conformal symmetry we found in the previous section, we formulated a model which has a much nicer sub-dimensional conformal symmetry. We still consider a complex scalar field $\phi$ in $2+1$ dimensions, only with a modified spatial derivative term given by the Lagrangian 
\begin{equation} \label{eq:2_2_model}
    \mathcal{L}  = \left|\partial_t \phi\right|^2 + \phi^* \partial_x \partial_y \phi.
\end{equation}
Unlike the previous model which had a discrete $D_4$ symmetry of the square lattice, this model only has the discrete $\mathbb{Z}_2 \times \mathbb{Z}_2$ symmetry of a rectangular lattice. The two $\mathbb{Z}_2$ factors act by exchanging $x$ and $y$ and by taking $(x,y) \rightarrow -(x,y)$ respectively. Finally, we note that there is an apparent instability in this theory as the spatial derivatives are not bounded from below, however as this is a quadratic model we can still formally study it. We will further discuss this instability, and how one should quantize this model below.

As before we can expand $\phi$ in terms of Fourier modes in the $y$ direction according to \eqref{eq:phi_of_ky}, resulting in the Lagrangian
\begin{equation}
    \mathcal{L}  = \left|\partial_t \phi_{k_y}\right|^2 + i k_y \phi_{k_y}^* \partial_x \phi_{k_y}.
\end{equation}
This Lagrangian describes a free non-relativistic particle, only with the roles of space and time interchanged. Thus this theory has a Schr\"odinger $1+1$ conformal symmetry \cite{Niederer:1972zz,Hagen:1972pd,1994JSP....75.1023H,Nishida:2007pj}. The symmetry consists of a $SL(2,\mathbb{R})$ of spatial reparameterizations which also acts on the time coordinate, along with translations in time and a Galilean boost. As $\phi$ has a scaling dimension of $1/2$ under the scaling symmetry, it too must transform under the $SL(2)$ portion of this symmetry. 

The $SL(2)$ portion of the transformation is
\begin{equation}
    x' = \frac{ax +b}{cx+d}, \qquad t' = \frac{t}{cx+d}, \qquad
    \phi_{k_y}' = (cx+d)^{1/2} e^{ik_y \frac{c t^2}{4(cx+d)}}\phi_{k_y},
    \qquad \qquad ad-bc =1,
\end{equation}
while the Galilean boost transforms the coordinates and field as
\begin{equation}
    t' = t+\alpha x, \qquad x' = x, \qquad
    \phi_{k_y}' = e^{ik_y \left(\frac{\alpha t}{2}+ \frac{\alpha^2 x}{4} \right)}\phi_{k_y}.
\end{equation}
Surprisingly, as the phase shift of $\phi_{k_y}$ is proportional to $k_y$, these symmetries are also a local transformation in the original coordinates. These transformations in the original coordinates are 
\begin{equation} \label{eq:sl2sym_x_coordinates}
    x' = \frac{ax +b}{cx+d}, \qquad t' = \frac{t}{cx+d}, \qquad
    y' = y + \frac{ct^2}{4(cx+d)}, \qquad \phi' = (cx+d)^{1/2} \phi,
    \qquad ad-bc =1,
\end{equation}
and
\begin{equation}\label{eq:boost_x_coordinates}
    t' = t+\alpha x, \qquad x' = x, 
    \qquad y' = y - \frac{a}{2} t - \frac{a^2}{4} x, 
    \qquad \phi' = \phi.
\end{equation}

This model also has a second Galilean conformal symmetry which can be seen when expanding $\phi$ in terms of momentum modes along the $x$ direction. The coordinate transformations of this symmetry is similar to \eqref{eq:sl2sym_x_coordinates} and \eqref{eq:boost_x_coordinates}, only with $x$ and $y$ interchanged. Both these symmetries also contain the same time translational symmetry.

To understand how these two independent symmetries play with each other, we look at their algebras. These symmetries are generated by the operators 
\begin{equation}
    \begin{aligned}
    &P_x = \partial_x, \qquad
    &&  D_x = 2x \partial_x +t \partial_t + \Delta, \qquad 
    && C_x = x^2 \partial_x + x t \partial_t -\frac{t^2}{2}\partial_y + x \Delta,\\
    & P_y = \partial_y, 
    && D_y = 2y \partial_y +t \partial_t + \Delta ,
    && C_y = y^2 \partial_y + y t \partial_t -\frac{t^2}{2}\partial_x + y \Delta,\\
    &  H = \partial_t, 
    && K_x = x \partial_t - t \partial_y , 
    && K_y = y \partial_t - t \partial_x,
    \end{aligned}
\end{equation}
where $\Delta$ is the scaling operator which acts on $\phi$ by $[\Delta,\phi] = -\frac{1}{2} \phi$. 

The $P$ operators generate the spatial translations, $H$ generates time translations and is shared by both conformal algebras, the $D$'s generate the scaling transformations, the $K$'s generate the Galilean boosts, and the $C$'s generate the special conformal transformations. The operators $P,D,$ and $C$ form the $\mathfrak{sl}(2)$ sub-algebra.

For these operators to form a closed algebra we need to add the operator
\begin{equation}
    C_0 = tx\partial_x +ty\partial_y + \left( \frac{t^2}{2} -xy \right) \partial_t  + t\Delta .
\end{equation}
Then these ten operators form a closed algebra with the commutation relations
\begin{equation}
    \begin{aligned}
    &[D_i,P_j] = -2\delta_{ij} P_i,
    &&[D_i,C_j] = 2\delta_{ij} C_i,
    &&[P_i,C_j] = \delta_{ij} D_i, \\
    & [H,P_i] = 0, 
    && [D_i,H] = -H,
    && [H,C_i] = K_i,\\
    & [K_i,H] = \sum_{j=x,y}\left(1-\delta_{ij}\right)P_j,
    && [P_i,K_j] = \delta_{ij} H,
    && [D_i, K_j] = \left(2 \delta_{ij}-1\right) K_j ,\\
    & [K_x,K_y] = \frac{1}{2}\left(D_y-D_x\right),
    && [C_i,K_j] = \left(1- \delta_{ij}\right) C_0,
    && [D_i,C_0] = C_0\\
    & [H,C_0] = \frac{1}{2}\left(D_y+D_x\right),
    && [P_i,C_0] = \sum_{j=x,y}\left(\delta_{ij}-1\right)K_j
    && [K_i,C_0] = C_i,
    \end{aligned}
\end{equation}
and all other nonstated commutators vanishing. From these commutations we immediately see that this is a simple lie algebra, and so must be one of the real forms of $\mathfrak{sp}(4,\mathbb{C}) = \mathfrak{so}(5,\mathbb{C})$. In fact it is already in the standard basis of $\mathfrak{sp}(4,\mathbb{C})$ (see for example \cite{hall2003lie} pg. 193), and so the real form is just $\mathfrak{sp}(4,\mathbb{R}) = \mathfrak{so}(3,2)$.

It seems surprising that we started from a non-Lorentz invariant theory in $3$ dimensions and found that the full conformal space-time symmetry is the same as that of a relativistic conformal field theory in $2+1$ dimensions. This however is due to the fact that the Lagrangian \eqref{eq:2_2_model} really describes a free boson in $2+1$ dimensions in the light cone coordinates \cite{Chang1973,Heinzl2000}. In this case the sub-dimensional Schr\"odinger symmetries conspire to generate the full 3 dimensional conformal symmetry of the model.

\section{The $(1,2)$ model} \label{sec:1_2_model}

As fractons are typically constructed as lattice models, there is no reason to assume they should be described by a relativistic field theory. Indeed in the previous section we studied a theory that had sub-dimensional Schr\"odinger symmetry, the non-relativistic analog of conformal symmetry. Thus in the next two sections we study analogous systems with an explicit non-relativistic time derivative term. 

The first model we consider is the non-relativistic version of the previous model, described by the Lagrangian
\begin{equation}
    \mathcal{L} = i \phi^* \partial_t \phi + \phi^* \partial_x \partial_y \phi.
\end{equation}
If we expand this in terms of momentum modes in the $y$ direction (or equivalently in the $x$ direction) we get a theory that looks like a free relativistic fermion in $1+1$ dimensions, or at least its holomorphic component. Thus this system has two $SL(2)$ sub-dimensional conformal symmetries.

To combine these two independent symmetries into a local conformal symmetry algebra we can use the same trick as before, and move to the relative coordinates
\begin{equation}
    u = \frac{x+y}{\sqrt{2}}, \qquad \qquad v = \frac{x-y}{\sqrt{2}}.
\end{equation}

Note that the spatial derivative term transforms as
\begin{equation}
    \partial_x\partial_y = \frac{1}{2} \left(\partial_u^2-\partial_v^2 \right),
\end{equation}
so the Lagrangian is really (up to total derivatives)
\begin{equation}
    \mathcal{L} = i \phi^* \partial_t \phi + \frac{1}{2}\left(\partial_v\phi^*\partial_v \phi -\partial_u\phi^*\partial_u \phi  \right),
\end{equation}
which is that of a free non-relativistic field in 2 dimensions with signature $(1,1)$. This theory has a slightly modified Schr\"odinger symmetry due to the $SO(1,1)$ spatial symmetry, rather than an $SO(2)$. Nevertheless, the symmetry is still well defined and generated by the central element $N$ (which generates the $U(1)$ phase symmetry of $\phi$), as well as the operators
\begin{equation}
    \begin{aligned}
    &H = \partial_t, && P_u = \partial_u, && P_v = \partial_v,\\
    &M = u\partial_u + v\partial_v, && D = 2t\partial_t+u\partial_u + v\partial_v + \Delta, &&K_u = t\partial_u-Nu\\
    &K_v = t \partial_v + Nv, \qquad && C = t^2 \partial_t +tu\partial_u +tv\partial_v+t\Delta - \frac{N}{2}\left(u^2-v^2\right).\qquad
    \end{aligned}
\end{equation}
We refer the readers to \cite{Hagen:1972pd} for a complete derivation of this symmetry algebra and the symmetry transformations.

We can translate this back to the original coordinates, and with some restructuring we find that the symmetry generators are
\begin{equation} \label{eq:symmetry_algebra_1_2}
    \begin{aligned}
    &H = \partial_t, && P_x = \partial_x, && P_y = \partial_y,\\
    &D_x = t\partial_t + x\partial_x+\frac{\Delta}{2},\qquad && D_y = t\partial_t + y\partial_y+\frac{\Delta}{2}, &&K_x = t\partial_x+Ny\\
    &K_y = t \partial_y + Nx,  && C = t^2 \partial_t +tx\partial_x +ty\partial_y+t\Delta + Nxy.\qquad
    \end{aligned}
\end{equation}

The actual symmetry transformations can also be constructed in the same manner, as their form is known for a free field in 2 spatial dimensions \cite{Hagen:1972pd}, and the Wick rotation of $SO(2)$ to $SO(1,1)$ is straightforward. The translations in space and scaling transformations are trivial to deduce. The Galilean boosts are given by the transformations
\begin{equation}
    x' = x+ct,\qquad \phi' = e^{icy}\phi, \qquad \text{ and } \qquad
    y' = y+ct,\qquad \phi' = e^{icx}\phi.
\end{equation}
Finally the transformations of the temporal $SL(2)$ symmetry are
\begin{equation}
t' = \frac{at+b}{ct+d}, \qquad x' = \frac{x}{ct+d}, 
\qquad y' = \frac{y}{ct+d}, \qquad \phi' = (ct+d)e^{i\frac{cxy}{ct+d}}\phi,
\qquad \qquad ad-bc = 1.
\end{equation}

This demonstrates a second example of a model where the two sub-dimensional conformal symmetries combine into a coherent symmetry algebra. In the previous example we had two non-relativistic sub-dimensional conformal symmetries generating a relativistic conformal symmetry, while in this model we had two relativistic sub-dimensional conformal symmetries generating a non-relativistic conformal symmetry.

We can use this toy model to build a Lagrangian that is similar to the $(2,4)$ model, only with a well defined conformal symmetry. Consider the Lagrangian
\begin{equation}
\label{eq:extension}
    \mathcal{L} = \left|\left(i\partial_t+\partial_x\partial_y\right)\phi\right|^2.
\end{equation}
This Lagrangian has the same symmetry algebra given by \eqref{eq:symmetry_algebra_1_2}, only now the dimension of $\phi$ is zero. Furthermore, when restricting $\phi$ to be real we obtain the same Lagrangian from \cite{Seiberg:2020bhn}\footnote{After Wick rotating time.}, so in a sense this is a different complexified version of the same model.

\section{The $(1,4)$ model} \label{sec:1_4_model}

We shall now discuss the last example we have of a sub-dimensional conformal symmetry, the non-relativistic version of the $(2,4)$ model. This model is described by the Lagrangian
\begin{equation}
    \mathcal{L} = i \phi^*\partial_t\phi + \left| \partial_x \partial_y \phi \right|^2.
\end{equation}

This is a surprisingly simple example of a model with fracton dynamics that is straightforward to study, as it is the second quantization of the quantum mechanical system with the Hamiltonian
\begin{equation}
    H = p_x^2 p_y^2.
\end{equation}
The spectrum of this model depends on the spectrum on the momentum operators $p_x$ and $p_y$. If they are quantized, say by taking space to be periodic, then the energy levels are 
\begin{equation}
    E_{m,n} \propto m^2 n^2.
\end{equation}
This implies the model has an infinite number of ground states given by $m=0$ or $n=0$, which correspond to states that have momentum in only one of the directions. We can put this system on a periodic lattice to regularize this infinity, when the ground state degeneracy becomes $2N_x+2N_y +1$, where $N_x$ and $N_y$ are the number of lattice points in the $x$ and $y$ directions. The divergence of the number of ground states as a function of system length, rather than volume, is a common theme among many fracton systems \cite{Nandkishore_2019,Pretko_2020}.\footnote{Here we should take $\phi$ to be a fermionic field so that the log of the ground state degeneracy goes like $2N_x+2N_y +1$, see appendix \ref{app:spectrum} for more details.}  We note that there are no quantum effects that lift the ground state degeneracy such as the ones found in \cite{Seiberg:2020bhn}.We present a more detailed analysis of the spectrum of this theory, as well as a comparison to the $(2,4)$ model and to \cite{Seiberg:2020bhn} in appendix \ref{app:spectrum}.

Such a large number of ground states with arbitrary momentum should make us doubt that this model accurately describes the low energies states with high momenta, and subsequently that it can describe the low energy behavior of physical systems. However, it does have the global dipole symmetry, similar to the one in \cite{Seiberg:2020bhn}, which may be imposed on the system to keep the ground state degeneracy intact.

Nevertheless, a better approach would be to add a small term to the Hamiltonian which breaks the ground state degeneracy. The simplest term which does this, and also preserves the discrete lattice symmetry $x \ra y$, $x \ra -x$, is the harmonic term
\begin{equation} \label{eq:harmonic_deformation}
    \delta H = \epsilon \left( p_x^2 + p_y^2 \right).
\end{equation}
This term breaks the ground state degeneracy and creates a hierarchy of states, the low energy excitations can have non-zero momentum in either the $x$ or the $y$ direction. To reach a state with momentum in both directions one needs to overcome a large energy gap. This dynamics is also similar to other theories of fractons \cite{Nandkishore_2019,Pretko_2020}.

To find the symmetry algebra, we expand the field in terms of momentum modes in one of the primary axes, say the $y$ direction. Then the theory becomes a free field in the remaining direction with a mass $m_x \propto k_y^{-2}$. Thus this theory has a sub-dimensional non-relativistic conformal symmetry, or Schr\"odinger symmetry \cite{Hagen:1972pd}. The $x$ sub-dimensional algebra is generated by the central element $N$ and the operators
\begin{equation}
    \begin{aligned}
        &H = \partial_t, && P_x = \partial_x, 
        && D_x = 2t\partial_t + x \partial_x + \Delta,\\
        & K_x = t\partial_x + \frac{x}{2k_y^2} N , \qquad
        && C_x = t^2\partial_t + tx\partial_x + t\Delta + \frac{x^2}{4k_y^2} N. \qquad
    \end{aligned}
\end{equation}
The $y$ sub-dimensional algebra is obtained by taking $x\mapsto y$. The commutation relations between the two algebras can be found by noting that $x$ and $k_x$ are conjugate variables, as are $y$ and $k_y$. The Galilean and special conformal transformations don't commute between the two algebras, and generate a large symmetry algebra that doesn't appear to close on itself, as happened in the $(2,4)$ model. 

The novelty here is that the relevant deformation \eqref{eq:harmonic_deformation} does not break the sub-dimensional conformal symmetry, but rather it deforms it to
\begin{equation}
    \begin{aligned}
        &H = \partial_t, 
        && D_x = 2t\partial_t + x \partial_x + \Delta+2\epsilon k_y^2 t N,
        & K_x = t\partial_x + \frac{x}{2(k_y^2+\epsilon)} N , \\
        & P_x = \partial_x, \quad
        && C_x = t^2\partial_t + tx\partial_x + t\Delta + \left(\frac{x^2}{4(k_y^2+\epsilon)}+\epsilon k_y^2t^2\right) N. \quad
    \end{aligned}
\end{equation}
This is in contrast to the $(2,4)$ model where this deformation breaks the sub-dimensional conformal symmetry. We note that this deformation does break the global dipole symmetry $\phi \mapsto \phi + c(x) + g(y)$ as in \cite{Seiberg:2020bhn}. This provides us with an example of a continuous family of theories which have a subsystem conformal symmetry, but which lack a global dipole symmetry. 

In fact we can deform the sub-dimensional conformal algebra all the way to the standard Schr\"odinger algebra. Consider a massive non-relativistic particle with an additional four derivative term
\begin{equation}
    \mathcal{L} = i \phi^*\partial_t\phi 
    + \frac{1}{2m} \left(\left|\partial_x\phi \right|^2
    + \left|\partial_y\phi \right|^2\right)
    + a \left| \partial_x \partial_y \phi \right|^2.
\end{equation}
This theory has a sub-dimensional conformal algebra for every value of the mass $m$ and parameter $a$. The generators of the symmetry in the $x$ direction take the form
\begin{equation}
    \begin{aligned}
        &H = \partial_t, 
        && D_x = 2t\partial_t + x \partial_x + \Delta+ \frac{1}{m} k_y^2 t N, \qquad \qquad  \qquad
        K_x = t\partial_x + \frac{mx}{(2mak_y^2+1)} N , \\
        & P_x = \partial_x, \quad
        && C_x = t^2\partial_t + tx\partial_x + t\Delta + \left(\frac{ m x^2}{2(2ma k_y^2+1)}+\frac{1}{2m} k_y^2t^2\right) N, \quad
    \end{aligned}
\end{equation}
and there is an additional set of generators for the symmetry in the $y$ direction. These two sets of generators do not commute unless $a=0$, when the algebra reduces to the standard Schr\"odinger algebra. We recover the original sub-dimensional conformal symmetry algebra in the limit $m\ra \infty$. 

The fact that the sub-dimensional conformal symmetry can be reached via a continuous deformations of the standard Schr\"odinger algebra may allow us to construct interacting theories with a sub-dimensional conformal algebra. One can imagine starting with an interacting non-relativistic CFT in $2+1$ dimensions, say Anyons in two spacial dimensions \cite{Nishida:2007pj}, and then continuously deform the theory (and the algebra) to realize an interacting theory with a sub-dimensional conformal algebra.

\section{Summary and Future Directions} \label{sec:summary}
We have presented four interesting examples of systems with sub-dimensional conformal symmetry in $2+1$ dimensions. Clearly there are many interesting future directions that follow from this work.

One obvious question is to ask to what extent our construction extends to other dimensions. While we focused on the 2+1 dimensional case, theories with subsystem conformal symmetry should also be easy to construct in other dimensions. One theory that can be written down in general $D$ spatial dimensions is based on a spatial gradient term in the Lagrangian of the form \cite{You_2020,Gorantla:2020xap}
\begin{equation}
    {\cal L} =\ldots +  (\partial_{x^1} \ldots \partial_{x^D} \phi )^2
\end{equation}
For every spatial direction $i$ this has a rescaling symmetry
\begin{equation}
D_i: \quad \quad  x^i \rightarrow \lambda x^i, \quad
t \rightarrow \lambda^z t
\end{equation}
with $z=1,2$ depending on whether we add a relativistic or non-relativistic time derivative term. These scale symmetries can easily be extended to conformal symmetries by the methods described in this work.

Beyond constructing other models, another obvious next step is to work out the consequences of the symmetries we uncovered. To what extent are correlation function constrained by the subsystem conformal symmetries? 

Last but not least, probably the most important question that needs to be addressed is how to construct interesting interacting field theories with the symmetries outlined in this work. Just like in \cite{Seiberg:2020bhn} our focus has been entirely on free field theories. Clearly we have just scratched the surface in our understanding of this kind of quantum field theory.

\appendix

\section{Spectral analysis and comparison of the $(1,4)$ and $(2,4)$ models} \label{app:spectrum}

Unlike in \cite{Seiberg:2020bhn} and the $(2,4)$ model, the ground state degeneracy of the $(1,4)$ model is exact, and cannot be lifted by quantum effects. We shall show this by comparing the spectrum of both models. For the $(1,4)$ model $\phi$ and $\phi^\dagger$ are conjugate operators that satisfy the canonical commutation relations
\begin{equation}
    \left[ \phi^\dagger(x),\phi(x') \right]_{\pm} = \delta^{(2)}(x-x') ,
\end{equation}
where we can take $\phi$ to be either a fermionic or a bosonic field. The Hamiltonian of the model reads
\begin{equation}
    H = \int dx ~dy~\partial_x \partial_y \phi^\dagger \partial_x \partial_y \phi.
\end{equation}
We can then move to the momentum mode variables by taking space to be compact and normalized, 
\begin{equation}
    \phi_{m,n} = \int dx~dy~ e^{i(mx+ny)} \phi(x,y),\qquad \qquad
    \phi_{m,n}^\dagger = \int dx~dy~ e^{i(mx+ny)} \phi(x,y)^\dagger,
\end{equation}
which satisfy the canonical commutation relations
\begin{equation}
    \left[ \phi^\dagger_{m,n}, \phi_{m'n'}\right]_{\pm} = \delta_{m,m'}\delta_{n,n'} .
\end{equation}
The Hamiltonian in terms of these variables is simply a sum of harmonic oscillators
\begin{equation}
    H = \sum_{m,n} m^2 n^2 \phi^\dagger_{m,n} \phi_{m,n}.
\end{equation}

We see that the ground states are really just harmonic oscillators with zero frequency, and so have zero energy for all occupancy numbers. In this case the ground state degeneracy is not lifted, and cannot be lifted even by taking $\phi$ to be compact.

To count the number of zero frequency modes we need to regularize this theory. The simplest regularization is by introducing cutoffs $N_x$ and $N_y$ for $m$ and $n$, which is equivalent to placing the theory on a lattice with lattice spacing $l_x \sim 1/N_x$, $l_y \sim 1/N_y$. The number of zero frequency modes with $n=0$ is $2N_x + 1$, as $|m|\leq N_x$, and similarly there are $2N_y+1$ zero modes with $m=0$. One of these modes is shared, the $m=n=0$ mode, giving us $2N_x+2n_y+1$ zero frequency modes. If $\phi$ is taken to be fermionic then the number of zero modes is $2^{2N_x+2n_y+1}$, and the log of the ground state degeneracy goes like $2N_x+2n_y+1$.

We note that this model is both gapless in the infinite volume limit and has an extensive number of exact ground states. This is peculiar as typically in gapless systems (say in conventional CFTs) a finite volume lifts the ground state degeneracy. Though calculating the ground state degeneracy of a gapless system may seem ill-defined as there are many states with vanishing energy, the number of exact ground states when placing the system in a finite volume is a well defined and measurable quantity.

We should contrast this with the relativistic $(2,4)$ model, where $\partial_t \phi^\dagger$ is the conjugate variable of $\phi$, and not $\phi^\dagger$. To understand the spectrum we first expand $\phi$ and $\phi^\dagger$ in terms of two real scalar fields, and then taking space to compact and normalized as above, the Hamiltonian in terms of the momentum modes will take the form
\begin{equation}
    H = \sum_{i=1,2} \sum_{m,n} \left( \pi_{i,m,n}^2 + m^2n^2 \phi_{i,m,n}^2\right),
\end{equation}
which is a sum of harmonic oscillators for $m,n\neq 0$, but a free theory for the modes where $m=0$ or $n=0$. The Hamiltonian for these modes is 
\begin{equation}
    H_{\text{"zero" modes}} = \sum_{i=1,2}\left( \sum_{m\neq 0}\pi_{i,m,0}^2+\sum_{n \neq 0} \pi_{i,0,n}^2 + \pi_{i,0,0}^2\right),
\end{equation}
which is a sum of free theories. For these modes the spectrum depends on the spectrum of the momentum operators $\pi_{i}$. If, as in \cite{Seiberg:2020bhn}, we take $\phi$ to be periodic then the spectrum of the $\pi$'s is quantized and this lifts their minimal energy except for the $(0,0)$ mode. This quantum effect would not happen if we didn't take $\phi$ to be periodic. These effects do not, and really cannot, happen in the non-relativistic models as the Hamiltonian is really a second quantized version of the quantum mechanical model.

\section*{Acknowledgments}
We would like to thank Nati Seiberg and Shu-Heng Shao and the anonymous reviewer for useful comments on the manuscript.
The work of AK was supported, in part, by the U.S.~Department of Energy under Grant No.~DE-SC0011637 and by a grant from the Simons Foundation (Grant 651440, AK).

\bibliography{conformalfractons.bib}

\bibliographystyle{JHEP}

\end{document}